\documentstyle[aps,psfig,eqsecnum,preprint]{revtex}
\begin{document}
\tighten
\draft
\preprint{{\small IFIC--00--43}}
\title{Diagonalization of the strongly coupled 
lattice QCD Hamiltonian}
\author{Yasuo Umino} 
\address{Departamento de Matematica Aplicada\\
Escuela Tecnica Superior de Ingenieros Industriales\\
Universidad Politecnica de Val\`encia\\
Val\`encia, Spain\\
and\\
Instituto de F\'{\i}sica Corpuscular -- C.S.I.C. \\
Departamento de F\'{\i}sica Te\`orica, Universitat de Val\`encia \\
E--46100 Burjassot, Val\`encia, Spain}
\date{\today}
\maketitle
\begin{abstract}
We construct a solution to the equation of motion of Hamiltonian lattice QCD
in the strong coupling limit using Wilson fermions which exactly diagonalizes
the Hamiltonian to second order in the field operators. This solution obeys the
free lattice Dirac equation with a dynamical mass which is identified with
the gap. The equation determining this gap is derived and it is found that the 
dynamical quark mass is a constant to lowest order in $N_{\rm c}$ but becomes momentum 
dependent once ${\cal O}(1/N_{\rm c})$ corrections are taken into account. We 
interpret our solution within the framework of the N--quantum 
approach to quantum field theory and discuss how our formalism may be
systematically extended to study bound states at finite temperature and chemical 
potential.
\vskip 0.25cm
\noindent Appeared in Physics Letters B 492 (2000) 385
\end{abstract}

\vfill\eject

\section{Introduction}
\label{intro}
The strong coupling approximation has played an important role in the development of
QCD lattice gauge theory from its very inception. In the renowned paper by
Wilson \cite{wil74} this approximation was invoked to demonstrate quark
confinement on the Euclidean space--time lattice. Soon thereafter Kogut and
Susskind \cite{kog75} formulated the Hamiltonian lattice gauge
theory and concluded that in the strong coupling limit the quark dynamics is
best described by a collection of non--Abelian electric flux tubes with quarks
attached at their ends. This was followed by the work of Baluni and
Willemsen who used a variant of the Kogut--Susskind formalism to demonstrate 
quantitatively that dynamical chiral symmetry breaking indeed takes places in 
lattice QCD at strong coupling \cite{bal76}. Finally, calculations by Kogut, 
Pearson and Shigemitsu \cite{kog79} and by Creutz \cite{cre79} suggesting the absence 
of a phase transition between the strong and weak coupling regimes of QCD
motivated numerous studies in this subject which continues to date. 

The Hamiltonian formulation of strong coupling QCD using Wilson fermions was first
studied in detail by Smit \cite{smi80} who derived an effective Hamiltonian
using the $1/N_{\rm c}$ approximation and applied the spin wave theory of magnetism
to determine the vacuum energy density and the excitation spectrum. In the
absence of the current quark mass and the Wilson term the theory posseses a
$U(4N_{\rm f})$ symmetry. Smit has shown how this symmetry is spontaneously broken 
down to $U(2N_{\rm f})\otimes U(2N_{\rm f})$ accompanied by the appearance of 
$8N_{\rm f}^2$ Goldstone bosons. A finite current quark mass also breaks the
original $U(4N_{\rm f})$ symmetry explicitly down to 
$U(2N_{\rm f})\otimes U(2N_{\rm f})$. Introduction of the Wilson term explicitly 
breaks the
latter symmetry further down to $U(N_{\rm f})$ solving the fermion doubling problem. 
In addition Smit has proposed how quantum corrections appearing at 
${\cal O}(1/N_{\rm c})$ may be used to resolve the $U_A(1)$ problem. 
Le~Yaouanc, Oliver, P\`{e}ne and Raynal \cite{ley86} later reexamined 
Smit's work by applying the Bogoliubov--Valatin variational method to the same effective 
Hamiltonian and found that the vacuum is not chirally degenerate and
concluded that the massless pseudoscalar boson found by Smit is not a
Goldstone boson.  

In this letter we report on a new approach to study strongly coupled Hamiltonian
lattice QCD using Wilson fermions. Our approach differs from those adopted in \cite{smi80}
and \cite{ley86} in that we explictly construct a solution to the field equations
of motion derived from the Hamiltonian. This solution exactly diagonalizes the
Hamiltonian to second order in field operators and obeys the free lattice
Dirac equation where the mass plays the role of the gap. Using the
equation of motion we derive the gap equation for the dynamical quark mass to
${\cal O}(1/N_{\rm c})$. Our results for the dynamical quark mass is
qualitatively different from the one found in \cite{ley86}. 

With our solution the interpretation of the
chiral condensate being the order parameter becomes transparent as well as the
observation that the broken symmetry phase is the energetically favored phase.
In addition, our approach admits to a systematic extension to the study of 
bound states both in free space and at finite temperature and chemical potential. 
This is accomplished by interpreting our solution within the context of the 
N--quantum approach to quantum field theory \cite{gre65,gre94a} which we
shall discuss in the concluding section.

\section{Non--interacting Wilson Fermions}
\label{wilson}

We begin with a brief review of the free lattice Dirac Hamiltonian with the Wilson
term. Although elementary, we shall see that the results presented here will play 
crucial roles in the construction of a solution of the strongly coupled QCD. We adopt 
the notation of Smit \cite{smi80} with unit lattice spacing $(a = 1)$ and temporarily 
suppress color and flavor indices. Then the free lattice Dirac Hamiltonian is
\begin{eqnarray}
H^0
& = &
\frac{1}{2i} \sum_{\vec{x},l}  
\left[ \Psi^\dagger(\vec{x}\,) \gamma_0\gamma_l \Psi(\vec{x}+\hat{n}_l)
- \Psi^\dagger(\vec{x}+\hat{n}_l) \gamma_0\gamma_l \Psi(\vec{x}\,) \right]
+ M \sum_{\vec{x}} \Psi^\dagger(\vec{x}\,) \gamma_0 \Psi(\vec{x}\,)
\nonumber\\
&   &
- \frac{r}{2} \sum_{\vec{x},l}
\left[ \Psi^\dagger(\vec{x}\,) \gamma_0 \Psi(\vec{x}+\hat{n}_l)
+ \Psi^\dagger(\vec{x}+\hat{n}_l) \gamma_0 \Psi(\vec{x}\,) \right]
\end{eqnarray}
where the third term is the Wilson term with $0 \leq r \leq 1$. For $r = 0$ there is 
an eightfold fermion multiplicity which is removed when $r \neq 0$.

At each lattice site the free Dirac field in configuration space is given by
\begin{equation}
\Psi_\mu(t,\vec{x}\,) = 
\sum_{\vec{p}} \left[ b(\vec{p}\,)\xi_\mu(\vec{p}\,)
e^{-i(\omega(\vec{p}\,) t - \vec{p}\cdot\vec{x})}
+ d^\dagger(\vec{p}\,)\eta_\mu(\vec{p}\,)
e^{i(\omega(\vec{p}\,) t - \vec{p}\cdot\vec{x})} \right]
\end{equation}
with $\mu$ denoting the Dirac index. The excitation energy $\omega(\vec{p}\,)$
will be determined shortly. The annihilation operators $b$ and $d$
annihilate the non--interacting vacuum state $|0\rangle$. For our purpose 
it is not necessary to know the structure of the spinors $\xi$ and $\eta$. The
only assumption that we shall make is that the creation and annihilation 
operators obey the free fermion anti--commutation relations
\begin{equation}
\left[b^\dagger(\vec{p}\,), b(\vec{q}\,)\right]_+
=
\left[d^\dagger(\vec{p}\,), d(\vec{q}\,)\right]_+
=
\delta_{\vec{p},\vec{q}}
\label{eq:antic}
\end{equation}
Using this assumption we can recover the anti--commutation relations for 
the field operators
\begin{equation}
\left[\Psi_\mu(t,\vec{x}\,), \Psi^\dagger_\nu(t,\vec{y}\,)\right]_+
=
\delta_{\vec{x},\vec{y}}\;\delta_{\mu\nu}
\end{equation}
provided that $\xi$ and $\eta$ satisfy the relation
\begin{equation}
\xi_\mu(\vec{p}\,) \xi^\dagger_\nu(\vec{p}\,) +
\eta_\mu(-\vec{p}\,) \eta^\dagger_\nu(-\vec{p}\,)
=
\delta_{\mu\nu}
\label{eq:compl}
\end{equation}
We normalize the spinors by demanding that the number density is given by 
\begin{equation}
{\cal N} 
= 
\sum_{\vec{x}} :\Psi^\dagger(t,\vec{x}\,)\Psi(t,\vec{x}\,): 
\; = 
2 \sum_{\vec{p}}
\left( b^\dagger(\vec{p}\,) b(\vec{p}\,)
- d^\dagger(\vec{p}\,) d(\vec{p}\,)\right)
\label{eq:ndensity}
\end{equation}
where the symbol :    : denotes normal ordering and the factor of 2 accounts for 
the spin degrees of freedom. Eq.~(\ref{eq:ndensity}) fixes the normalizations 
of $\xi$ and $\eta$ to be  
\begin{eqnarray}
\xi_\mu(\vec{p}\,) \xi^\dagger_\mu(\vec{p}\,) 
& = & \eta_\mu(\vec{p}\,) \eta^\dagger_\mu(\vec{p}\,)
 =  2 
\label{eq:norm1}\\
\xi^\dagger_\mu(\vec{p}\,) \eta_\mu(-\vec{p}\,)
& = &
\eta^\dagger_\mu(\vec{p}\,) \xi_\mu(-\vec{p}\,)  
 =  0 
\label{eq:norm2}
\end{eqnarray}
which are consistent with Eq.~(\ref{eq:compl}).

We now go over into momentum space where we perform all our calculations. In
momentum space the charge conjugaton symmetric form of $H^0$ is \footnote{Our 
convention for the Fourier transform from configuration to momentum space is
$\Psi(\vec{x}\,) =  \sum_{\vec{p}}\Psi(\vec{p}\,)e^{\vec{p}\cdot\vec{x}}$
which implies that the volume $V$ is given by
$V = \sum_{\vec{x}} = \delta_{\vec{p},\vec{p}}.$} 
\begin{equation}
H^0 = 
\frac{1}{2} \sum_{\vec{p}} 
\Bigl( - \sum_{l} \sin(\vec{p}\cdot\hat{n}_l) \gamma_0\gamma_l
+ M(\vec{p}\,) \gamma_0 \Bigr)_{\mu\nu}
\biggl[ \Psi^\dagger_\mu(t,\vec{p}\,), \Psi_\nu(t,-\vec{p}\,) \biggr]_-
\label{eq:h0}
\end{equation}
where the momentum dependent mass term is given by 
\begin{equation}
M(\vec{p}\,) \equiv M - r \sum_l \cos(\vec{p}\cdot\hat{n}_l)
\label{eq:massp}
\end{equation}
The free Dirac field becomes 
\begin{equation}
\Psi_\mu(t,\vec{p}\,) = 
b(\vec{p}\,)\xi_\mu(\vec{p}\,)e^{-i\omega(\vec{p}\,) t}
+ d^\dagger(-\vec{p}\,)\eta_\mu(-\vec{p}\,)e^{+i\omega(\vec{p}\,) t}
\label{eq:psip} 
\end{equation}
which is used to derive the equation of motion corresponding to Eq.~(\ref{eq:h0})
\begin{eqnarray}
i \dot{\Psi}(t,\vec{p}\,)
& = &
:\left[\Psi(t,\vec{p}\,), H^0\right]_-:
\\
& = & 
\biggl( \sum_{l} \sin(\vec{p}\cdot\hat{n}_l) \gamma_0\gamma_l
+ M(\vec{p}\,) \gamma_0 \biggr) \Psi(t,\vec{p}\,)
\label{eq:eomh0}
\end{eqnarray}
From Eq.~(\ref{eq:eomh0}) one obtains the excitation energy
\begin{equation}
\omega(\vec{p}\,) =
\biggl( \sum_{l} \sin^2(\vec{p}\cdot\hat{n}_l) 
+ M^2(\vec{p}\,)\biggr)^{1/2}
\end{equation}
and the equations of motion for the $\xi$ and $\eta$ spinors
\begin{eqnarray}
\omega(\vec{p}\,)\; \xi(\vec{p}\,)
& = & 
\biggl( \sum_{l} \sin(\vec{p}\cdot\hat{n}_l) \gamma_0\gamma_l
+ M(\vec{p}\,) \gamma_0 \biggr) \xi(\vec{p}\,)
\label{eq:eomxi} \\
\omega(\vec{p}\,)\; \eta(-\vec{p}\,)
& = & 
- \biggl( \sum_{l} \sin(\vec{p}\cdot\hat{n}_l) \gamma_0\gamma_l
+ M(\vec{p}\,) \gamma_0 \biggr) \eta(-\vec{p}\,)
\label{eq:eometa}
\end{eqnarray}

When $r=0$ these equations of motion are relativistic near the eight corners
of the Brillouin zone denoted by
$\vec{\pi}_0 = (0, 0, 0)$,
$\vec{\pi}_{\rm x} = (\pi, 0, 0)$,
$\vec{\pi}_{\rm y} = (0, \pi, 0)$,
$\vec{\pi}_{\rm z} = (0, 0, \pi)$,
$\vec{\pi}_{\rm xy} = (\pi, \pi, 0)$,
$\vec{\pi}_{\rm xz} = (\pi, 0, \pi)$,
$\vec{\pi}_{\rm yz} = (0, \pi, \pi)$ and
$\vec{\pi}_{\rm xyz} = (\pi, \pi, \pi)$. The excitation energies at these values
of momenta are equal which corresponds to the eightfold multiplicity mentioned
above. This degeneracy is lifted when $r \neq 0$ due to the momentum
dependent mass term Eq.~(\ref{eq:massp}). Using the equations of motion for $\xi$
and $\eta$ it is a simple excercise to show that the off--diagonal Hamiltonian 
vanishes and that the vacuum energy is given
by $\langle 0|H^0|0 \rangle = -2 V\sum_{\vec{p}} \omega(\vec{p}\,)$.
Finally, we construct positive and negative energy projection 
operators $\Lambda^+(\vec{p}\,)$ and $\Lambda^-(\vec{p}\,)$ as follows
\begin{eqnarray}
\Lambda^+(\vec{p}\,)
& \equiv &
\xi(\vec{p}\,)\otimes\xi^\dagger(\vec{p}\,)
\nonumber \\
& = &
\frac{1}{2} \left[ 1 + 
\frac{1}{\omega(\vec{p}\,)} \sum_{l} \sin(\vec{p}\cdot\hat{n}_l) \gamma_0\gamma_l
+ \frac{M(\vec{p}\,)}{\omega(\vec{p}\,)} \gamma_0 \right]
\label{eq:projp} \\
\Lambda^-(\vec{p}\,)
& \equiv &
\eta(-\vec{p}\,)\otimes\eta^\dagger(-\vec{p}\,)
\nonumber \\
& = &
\frac{1}{2} \left[ 1 -
\frac{1}{\omega(\vec{p}\,)} \sum_l \sin(\vec{p}\cdot\hat{n}_l) \gamma_0\gamma_l
- \frac{M(\vec{p}\,)}{\omega(\vec{p}\,)} \gamma_0 \right]
\label{eq:projn} 
\end{eqnarray}
These projection operators will be used extensively below. Note that they 
obey the condition
\begin{equation}
\left[ \Lambda^+ (\vec{p}\,) + \Lambda^-(\vec{p}\,)\right]_{\alpha\beta}
= \delta_{\alpha\beta}
\end{equation}
as is required by Eq.~(\ref{eq:compl}). 

\section{Effective Hamiltonian and the Ansatz}

For the sake of comparison we use the same effective Hamiltonian obtained by Smit
in Eq.~(3.13) of \cite{smi80} which is also used in \cite{ley86}. The charge 
conjugation symmetric form of Smit's Hamiltonian in momentum space is
\begin{eqnarray}
H_{\rm eff}
& = &
\frac{1}{2} M_0 \sum_{\vec{p}_1,\vec{p}_2} \delta_{\vec{p}_1+\vec{p}_2,\vec{0}}
\left( \gamma_0\right)_{\mu\nu} 
\biggl[ \bigl(\Psi^\dagger_{a\alpha}\bigr)_\mu(\vec{p}_1\,),
\bigl(\Psi_{a\alpha}\bigr)_\nu(\vec{p}_2\,) \biggr]_-
\nonumber \\
&   & -\frac{K}{8N_{\rm c}} \sum_{\vec{p}_1,\ldots,\vec{p}_4 }\sum_{l}
\delta_{\vec{p}_1+\cdot\cdot\cdot+\vec{p}_4,\vec{0}}
\Biggl[ e^{i((\vec{p}_1+\vec{p}_2)\cdot \hat{n}_l)} 
+ e^{i((\vec{p}_3+\vec{p}_4)\cdot \hat{n}_l)} \Biggr]
\nonumber \\
&   &
\otimes \Biggl[
\left( \Sigma_l \right)_{\mu\nu} 
\bigl(\Psi^\dagger_{a\alpha}\bigr)_\mu(\vec{p}_1\,)  
\bigl( \Psi_{b\alpha}\bigr)_\nu(\vec{p}_2\,)
-
\left( \Sigma_l \right)^\dagger_{\mu\nu}
\bigl(\Psi_{a\alpha}\bigr)_\nu(\vec{p}_1\,)
\bigl( \Psi^\dagger_{b\alpha}\bigr)_\mu(\vec{p}_2\,) \Biggr]
\nonumber \\
&   &
\otimes \Biggl[
\left( \Sigma_l \right)^\dagger_{\gamma\delta} 
\bigl(\Psi^\dagger_{b\beta}\bigr)_\gamma(\vec{p}_3\,)  
\bigl(\Psi_{a\beta}\bigr)_\delta(\vec{p}_4\,)
-
\left( \Sigma_l \right)_{\gamma\delta}
\bigl(\Psi_{b\beta}\bigr)_\delta(\vec{p}_3\,)
\bigl( \Psi^\dagger_{a\beta}\bigr)_\gamma(\vec{p}_4\,) \Biggr]
\label{eq:heffp}
\end{eqnarray}
where $\Sigma_l \equiv -i(\gamma_0\gamma_l - ir\gamma_0)$. We denote color,
flavor and Dirac indices by $(ab)$, $(\alpha\beta)$ and
$(\mu\nu\gamma\delta)$, respectively. Summation
convention for repeated indices is implied. The three parameters in this
theory are the Wilson parameter $r$, the current quark mass $M_0$ and the effective 
coupling constant $K$ which behaves as $1/g^2$ with $g$ being the QCD coupling constant.

Our ansatz which diagonalizes Eq.~(\ref{eq:heffp}) to second order in the field
operators is deceptively simple. It is the solution of the free lattice
Dirac equation Eq.~(\ref{eq:eomh0}) with $r = 0$ and is given by Eq.~(\ref{eq:psip})
appropriately modified to include color and flavor degrees of freedom.\footnote{This
modification will of course affect the anti--commutation relation
Eq.~(\ref{eq:antic}).} However, the mass term $M(\vec{p}\,)$ in Eq.~(\ref{eq:eomh0})
is now the dynamical and {\em not} the current quark 
mass. We shall show below that this dynamical mass is the solution of the
gap equation. In addition, the annihilation operators $b$ and
$d$ in our ansatz now annihilate an interacting vacuum state $|{\cal G}\rangle$
and {\em not} $|0\rangle$. Since we shall be working only in the space of quantum
field operators there is no need to specify the structure of
$|{\cal G}\rangle$. Aside from these non--trivial modifications our ansatz obeys all 
the properties of the free Dirac field described above. For example, with this ansatz 
the chiral condensate is given by
\begin{eqnarray}
\langle {\cal G}|\bar{\Psi}\Psi|{\cal G} \rangle
& = &
\frac{N_{\rm c} N_f}{2} \sum_{\vec{p}}
\left[ {\rm Tr}\bigl(\gamma_0\Lambda^-(\vec{p}\,)\bigr)
- {\rm Tr}\bigl(\gamma_0\Lambda^+(\vec{p}\,)\bigr) \right]
\nonumber \\
& = &
-2 N_{\rm c} N_f \sum_{\vec{p}} \frac{M(\vec{p}\,)}{\omega(\vec{p}\,)}
\label{eq:chicon}
\end{eqnarray}
Hence the chiral condensate is determined by the dynamical quark mass 
and clearly plays the role of the order parameter. The basic idea of 
our approach is to use the equation of motion for $H_{\rm eff}$ to derive
and solve the gap equation for the dynamical mass. 

Using our ansatz we proceed to derive the vacuum energy density
and the second order off--diagonal Hamiltonian by renormal ordering 
$H_{\rm eff}$ with respect to $|{\cal G}\rangle$. The former is given by 
terms involving two contractions while the latter is derived by
retaining terms involving a single contraction. Our results for the vacuum energy 
density is
\begin{eqnarray}
\frac{1}{V} \langle {\cal G}|H_{\rm eff}|{\cal G} \rangle
& = &
-2 M_0 N_{\rm c} \sum_{\vec{p}} {\rm Tr}\left[\Lambda^+(\vec{p}\,)\gamma_0\right]
\nonumber \\
&   &
-K\sum_{\vec{p},\vec{q}} \sum_l 
\Lambda^+_{\nu\mu}(\vec{p}\,) \Lambda^+_{\delta\gamma}(\vec{q}\,)
\nonumber \\
&   &
\otimes\left\{ \cos\left( \vec{p}-\vec{q}\;\right)\cdot\hat{n}_l
\Biggl[
\bigl(\Sigma_l\bigr)_{\mu\nu}\bigl(\Sigma_l\bigr)^\dagger_{\gamma\delta}
+
\bigl(\Sigma_l\bigr)^\dagger_{\mu\nu}\bigl(\Sigma_l\bigr)_{\gamma\delta}
\Biggr]\right.
\nonumber \\
&   &
\:\:\:\:\:\:\:\:\:\:
\left.
+
\cos\left( \vec{p}+\vec{q}\;\right)\cdot\hat{n}_l
\Biggl[
\bigl(\Sigma_l\bigr)^\dagger_{\mu\nu}\bigl(\Sigma_l\bigr)^\dagger_{\gamma\delta}
+
\bigl(\Sigma_l\bigr)_{\mu\nu}\bigl(\Sigma_l\bigr)_{\gamma\delta}
\Biggr]\right\}
\nonumber \\
&   &
-\frac{1}{2}K\sum_{\vec{p},\vec{q}} \sum_l 
\left[ \Lambda^+_{\nu\gamma}(\vec{q}\,) - \delta_{\nu\gamma} \right]
\Lambda^+_{\delta\mu}(\vec{p}\,)
\nonumber \\
&   &
\otimes\left\{ N_{\rm c}
\Biggl[
\bigl(\Sigma_l\bigr)_{\mu\nu}\bigl(\Sigma_l\bigr)^\dagger_{\gamma\delta}
+
\bigl(\Sigma_l\bigr)^\dagger_{\mu\nu}\bigl(\Sigma_l\bigr)_{\gamma\delta}
\Biggr]\right.
\nonumber \\
&   &
\:\:\:\:\:\:\:\:\:\:
\left.
+
\cos\left( \vec{p}+\vec{q}\;\right)\cdot\hat{n}_l
\Biggl[
\bigl(\Sigma_l\bigr)^\dagger_{\mu\nu}\bigl(\Sigma_l\bigr)^\dagger_{\gamma\delta}
+
\bigl(\Sigma_l\bigr)_{\mu\nu}\bigl(\Sigma_l\bigr)_{\gamma\delta}
\Biggr]\right\}
\label{eq:ved}
\end{eqnarray}
while the second order off--diagonal Hamiltonian is found to be
\begin{eqnarray}
H_{\rm off}|{\cal G}\rangle
& = &
\Biggl\{ 
M_0 \sum_{\vec{p}} \bigl( \gamma_0 \bigr)_{\mu\nu}
\xi^\dagger_\mu(\vec{q}\,) \eta_\nu(-\vec{q}\,)
\nonumber \\
&   &
+ \frac{1}{N_{\rm c}}K\sum_{\vec{p},\vec{q}}\sum_l \Lambda^+_{\nu\mu}(\vec{q}\,)
\xi^\dagger_\gamma(\vec{q}\,) \eta_\delta(-\vec{q}\,)
\nonumber \\
&   &
\otimes\left[ \cos\left( \vec{p}-\vec{q}\;\right)\cdot\hat{n}_l
\Biggl(
\bigl(\Sigma_l\bigr)_{\mu\nu}\bigl(\Sigma_l\bigr)^\dagger_{\gamma\delta}
+
\bigl(\Sigma_l\bigr)^\dagger_{\mu\nu}\bigl(\Sigma_l\bigr)_{\gamma\delta}
\Biggr)\right.
\nonumber \\
&   &
\:\:\:\:\:\:\:\:\:\:
\left.
+
\cos\left( \vec{p}+\vec{q}\;\right)\cdot\hat{n}_l
\Biggl(
\bigl(\Sigma_l\bigr)^\dagger_{\mu\nu}\bigl(\Sigma_l\bigr)^\dagger_{\gamma\delta}
+
\bigl(\Sigma_l\bigr)_{\mu\nu}\bigl(\Sigma_l\bigr)_{\gamma\delta}
\Biggr)\right]
\nonumber \\
&   &
-\frac{1}{N_{\rm c}}\frac{K}{4} \sum_{\vec{p},\vec{q}} \sum_l 
\left[ 2\Lambda^+_{\nu\gamma}(\vec{p}\,) - \delta_{\nu\gamma} \right]
\xi^\dagger_\mu(\vec{q}\,) \eta_\delta(-\vec{q}\,)
\nonumber \\
&   &
\otimes\left[ N_{\rm c}
\Biggl(
\bigl(\Sigma_l\bigr)_{\mu\nu}\bigl(\Sigma_l\bigr)^\dagger_{\gamma\delta}
+
\bigl(\Sigma_l\bigr)^\dagger_{\mu\nu}\bigl(\Sigma_l\bigr)_{\gamma\delta}
\Biggr)\right.
\nonumber \\
&   &
\:\:\:\:\:\:\:\:\:\:
\left.
+
\cos\left( \vec{p}+\vec{q}\;\right)\cdot\hat{n}_l
\Biggl(
\bigl(\Sigma_l\bigr)^\dagger_{\mu\nu}\bigl(\Sigma_l\bigr)^\dagger_{\gamma\delta}
+
\bigl(\Sigma_l\bigr)_{\mu\nu}\bigl(\Sigma_l\bigr)_{\gamma\delta}
\Biggr)\right] \Biggr\}e^{i2\omega(\vec{q}\,) t}
\nonumber \\
&   &
\:\:\:\:\:\:\:\:\:\:
\otimes b^\dagger_{\alpha,a}(\vec{q}\,)d^\dagger_{\alpha,a}(-\vec{q}\,)
|{\cal G}\rangle
\label{eq:hoff2}
\end{eqnarray}
We see from Eq.~(\ref{eq:hoff2}) that the excitation spectrum of the effective 
Hamiltonian involves color singlet quark--anti--quark mesonic excitations coupled 
to zero total three momentum. 

\section{Equation of Motion and Diagonalization of $H_{\rm eff}$}

We shall now demonstrate that $H_{\rm off}$ vanishes exactly by exploiting the 
equation of motion for $H_{\rm eff}$. For this purpose it suffices to consider 
the equation of motion for the up quark field $u(t,\vec{q}\,)$ given by
\begin{eqnarray}
i \bigr(\dot{u}_a\bigl)_\mu(t,\vec{q}\,)
& = &
: \left[\bigr(u_a\bigl)_\mu(t,\vec{q}\,), H_{\rm eff} \right]_-:
\label{eq:eomheff}
\\
& = & 
\Biggl\{
M_0 \left(\gamma_0\right)_{\mu\delta}
\\
&   & 
+ \frac{1}{N_{\rm c}}K\sum_{\vec{p}}\sum_l \Lambda^+_{\nu\gamma}(\vec{p}\,)
\nonumber \\
&   &
\otimes\left[ \cos\left( \vec{p}-\vec{q}\;\right)\cdot\hat{n}_l
\Biggl(
\bigl(\Sigma_l\bigr)_{\gamma\nu}\bigl(\Sigma_l\bigr)^\dagger_{\mu\delta}
+
\bigl(\Sigma_l\bigr)^\dagger_{\mu\nu}\bigl(\Sigma_l\bigr)_{\gamma\delta}
\Biggr)\right.
\nonumber \\
&   &
\:\:\:\:\:\:\:\:\:\:
\left.
+
\cos\left( \vec{p}+\vec{q}\;\right)\cdot\hat{n}_l
\Biggl(
\bigl(\Sigma_l\bigr)^\dagger_{\gamma\nu}\bigl(\Sigma_l\bigr)^\dagger_{\mu\delta}
+
\bigl(\Sigma_l\bigr)_{\mu\nu}\bigl(\Sigma_l\bigr)_{\gamma\delta}
\Biggr)\right]
\nonumber \\
&   &
-\frac{1}{N_{\rm c}}\frac{K}{4} \sum_{\vec{p},\vec{q}} \sum_l 
\left[ 2\Lambda^+_{\nu\gamma}(\vec{p}\,) - \delta_{\nu\gamma} \right]
\nonumber \\
&   &
\otimes\left[ N_{\rm c}
\Biggl(
\bigl(\Sigma_l\bigr)_{\mu\nu}\bigl(\Sigma_l\bigr)^\dagger_{\gamma\delta}
+
\bigl(\Sigma_l\bigr)^\dagger_{\mu\nu}\bigl(\Sigma_l\bigr)_{\gamma\delta}
\Biggr)\right.
\nonumber \\
&   &
\:\:\:\:\:\:\:\:\:\:
\left.
+
\cos\left( \vec{p}+\vec{q}\;\right)\cdot\hat{n}_l
\Biggl(
\bigl(\Sigma_l\bigr)^\dagger_{\mu\nu}\bigl(\Sigma_l\bigr)^\dagger_{\gamma\delta}
+
\bigl(\Sigma_l\bigr)_{\mu\nu}\bigl(\Sigma_l\bigr)_{\gamma\delta}
\Biggr)\right] \Biggr\} \left(u_a\right)_\delta(t, \vec{q}\,)
\end{eqnarray}
The crucial step in diagonalizing $H_{\rm eff}$ is to recall that $u_a$ is
also the solution of the free lattice Dirac equation with $r = 0$ and therefore 
one can identify Eq.~(\ref{eq:eomheff}) with Eq.~(\ref{eq:eomh0}).
\begin{equation}
\Bigl[u_a, H_{\rm eff}\Bigr]_-
=
\Bigl[u_a, H^0_{r = 0}\Bigr]_-
\label{eq:crux}
\end{equation}
Then the second order off--diagonal Hamitonian vanishes due to the equation of 
motion for $\eta$  Eq.~(\ref{eq:eometa}) and the orthogonality condition 
Eq.~(\ref{eq:norm2}).
\begin{eqnarray}
H_{\rm off}|{\cal G}\rangle 
& = &
\xi^\dagger(\vec{q}\,)
\left( \sum_{l} \sin(\vec{q}\cdot\hat{n}_l) \gamma_0\gamma_l
+ M(\vec{q}\,) \gamma_0 \right) \eta(-\vec{q}\,) 
e^{i2\omega(\vec{q}\,) t} 
b^\dagger_{\alpha,a}(\vec{q}\,)d^\dagger_{\alpha,a}(-\vec{q}\,) |{\cal G}\rangle
\nonumber \\
& = &
0
\end{eqnarray}

We can also use Eq.~(\ref{eq:crux}) to simplify the vacuum energy density 
to the following form.
\begin{eqnarray}
\frac{1}{V} \langle {\cal G}|H_{\rm eff}|{\cal G}\rangle
& = &
-N_{\rm c} \sum_{\vec{q}} \left\{ 3K(1+r^2) 
+ 2\omega(\vec{q}\,)
+\frac{M(\vec{q}\,)}{\omega(\vec{q}\,)}M_0 \right\}
\label{eq:ved2}
\end{eqnarray}
Therefore, the difference in the vacuum energy densities between the phase with 
$M(\vec{q}\,) = 0$ and $M(\vec{q}\,) \neq 0$ is given by
\begin{eqnarray}
\Delta E 
& = &
\frac{1}{V}\langle {\cal G}|H_{\rm eff}|{\cal G} \rangle |_{M(\vec{q}\,) = 0}
- \frac{1}{V}\langle {\cal G}|H_{\rm eff}|{\cal G} \rangle |_{M(\vec{q}\,) \neq 0}
\nonumber \\
& = &
N_{\rm c} \sum_{\vec{q}} \left[ 
\frac{M(\vec{q}\,)}{\omega(\vec{q}\,)}M_0 + 
2 \left( \sum_l \sin^2(\vec{q}\cdot\hat{n}_l) + M^2(\vec{q}\,)\right)^{1/2}
- 2 \left( \sum_l \sin^2(\vec{q}\cdot\hat{n}_l)\right)^{1/2}
\right]
\nonumber \\
& > &
0
\end{eqnarray}
Hence the phase with a finite gap is the energetically favored phase. 
It remains to derive the gap equation and to solve it for the dynamical quark 
mass.

\section{The Gap Equation and the Dynamical Quark Mass}

The gap equation is obtained by using Eq.~(\ref{eq:crux}) and equating the
coefficients of the $\gamma_0$ operator. Using Eq.~(\ref{eq:projp}) we find 
\begin{eqnarray}
M(\vec{q}\,)
& = &
M_0 + \frac{3}{2}K(1 - r^2) \sum_{\vec{p}} \frac{M(\vec{p}\,)}{\omega(\vec{p}\,)}
\nonumber \\
&   &
+ \frac{1}{N_{\rm c}} K\sum_{\vec{p},l} \frac{M(\vec{p}\,)}{\omega(\vec{p}\,)}
\Biggl\{
8r^2  \cos(\vec{p}\cdot\hat{n}_l) \cos(\vec{q}\cdot\hat{n}_l)
-\frac{1}{2}(1+r^2) \cos(\vec{p}+\vec{q}\, )\cdot\hat{n}_l
\Biggr\}
\label{eq:gap} 
\end{eqnarray}
Thus the dynamical quark mass is determined by a three dimensional
non--linear integral equation and is momentum dependent. This
dependence comes from the ${\cal O}(1/N_{\rm c})$ correction to the 
gap equation.

The solution to the gap equation to lowest order in $N_{\rm c}$ is shown in
Figure~1a for the case of $M_0 = 0$ and $r = 0$. We see that the dynamical 
symmetry breaking takes place
only above a critical coupling constant of about $K_{\rm C} \approx 0.73$.
From Eq.~(\ref{eq:gap}) we see that $K_{\rm C}$ must
increase as $r$ is increased in order for the gap equation to have a solution and
this situation is shown in Figure~1b. In fact, the critical coupling constant
approaches infinity as $r \rightarrow 1$ and the assumption of strong
coupling (small $K$) breaks down in this limit. 

However, when $r = 1$ there is no ${\cal O}(N^0_{\rm c})$ contribution to 
Eq.~(\ref{eq:gap}) and the dynamical quark mass may be written as 
\begin{equation}
M(\vec{q}\,) = M_0 + B(K)\sum_l \cos(\vec{q}\cdot\hat{n}_l)
\label{eq:gapq}
\end{equation}
assuming that $M(-\vec{q}\,) = M(\vec{q}\,)$. In this case the solution to the gap 
equation is shown in Figure~2 again for the $M_0 = 0$ case. At this value of the Wilson
parameter we find solutions to the gap equation only above a value of 
$K_{\rm C} \approx 0.84$. However, the dynamical quark mass is one order of
magnitude less than that obtained with only the ${\cal O}(N^0_{\rm c})$ contribution.
This result suggests that the momentum dependence of the dynamical
quark mass may be treated as a correction to the dominant constant term.

Our results for the dynamical mass is qualitatively different to the one
obtained by Le~Yaouanc, Oliver, P\`{e}ne and Raynal in Eq.~(6.6) of \cite{ley86}.
Not only their dynamical mass is momentum independent, but the chiral
symmetry breaking takes place for all values of the coupling constant $K > 0$.
In addition, from Eq.~(\ref{eq:chicon}) we see that in the
$M_0 \rightarrow 0$ limit the chiral condensate is finite only above a certain
value of the critical coupling constant $K_{\rm C}$ for the examples
presented here. Above $K_{\rm C}$, the chiral condensate is not a constant but
a function of $K$. This situation is in contrast to the behaviour of the 
chiral condensate obtained in \cite{ley86} where it is always a finite constant
in the broken symmetry mode.

\section{Conclusion and Outlook}

In this work we have constructed a simple solution to Smit's effective Hamiltonian 
for the strongly coupled lattice QCD. Our solution obeys the free lattice Dirac 
equation with a dynamical quark mass. This mass is the 
solution of the gap equation which is determined from the equation of motion.
The creation and annihilation operators for quarks and anti--quarks obey the 
well--defined free fermion anti--commuation relations and annihilate an interacting 
vacuum state. Since quantites of interest are solely determined by the 
anti--commutation relations there is no need to specify the structure of
this state.

The solution presented here exactly diagonalizes the 
Hamiltonian to second order in the quark field operators. We 
find that the elementary excitations of the theory consist of quark--anti--quark color
singlet states coupled to zero total three momentum and that the phase with
broken chiral symmetry is the energetically favored phase. Morever in this phase
the gap of the theory is found to depend both on momentum and on the effective coupling
constant.

However, our solution is by no means complete. It may be systematically improved
to calculate the masses, widths and coupling constants of all the bound states
allowed by the Hamiltonian if our approach is to be interpreted within the context of 
the N--quantum approach (NQA) to quantum field theory \cite{gre65,gre94a} as follows. 
NQA is a method to solve the field equations of motion by expanding the Heisenberg 
fields in terms of asymptotic fields obeying the free field equations of motion. This
expansion, which varies from theory to theory, is known as the Haag expansion 
\cite{haa55}. 

The first term in the Haag expansion is simply a single free asymptotic
field. Previous applications of the NQA to various field theories have shown that
the use of this first order term alone can reproduce the known mean field results
not only in free space but also at finite temperature ($T$) and chemical
potential ($\mu$) \cite{gre94a,umi00a}. These additional variables are introduced by 
subjecting each of the field operators in the Haag expansion to thermal Bogoliubov
transformations just as in thermal field dynamics. The solution presented in this work 
can be interpreted as the first order Haag expansion of the Heisenberg quark fields 
in the effective strong coupling QCD Hamiltonian. Since the time coordinate is
not discreticized in the Hamiltonian (Kogut--Susskind) formulation of lattice gauge 
theory one can naturally introduce and work with the concept of an asymptotic field. 

We can improve our solution by taking the second order terms in the Haag
expansion into account. In our case each of these terms would be a product of
a fermionic (quark) and a bosonic field, the latter corresponding to the
various bound states allowed by the Hamiltonian. They are the elementary
excitations identified in this work. The coefficients of the second order
terms are interpreted as amplitudes for formations of these bound states.
Self--consistency equations for these bound state amplitudes are derived by 
using the equation of motion. These bound state amplitudes have the same
number of kinematical variables as in non--relativistic theories and are
independent of Bethe--Salpter amplitudes. The idea of the NQA is to solve
for these amplitudes and thus construct a solution to the field theory
beyond the mean field approximation.

The program described above have recently been carried out for the effective
instanton induced `t~Hooft interaction at finite $T$ and $\mu$ \cite{umi00b}.
In constructing a solution to the `t~Hooft model beyond the mean field theory the
masses, widths and coupling constants of the $\sigma$ and diquark states have been 
determined. It would be interesting to apply the second order NQA to the Hamiltonian 
considered in this work and to calculate the properties of bound states of the theory 
and to construct an improved solution to the strong coupling QCD Hamiltonian at any 
$T$ and $\mu$. This solution can then be
used to obtain a qualitative picture of the QCD equation of state. 

\acknowledgements
I thank M.--P.~Lombardo for suggesting to apply the NQA to strongly coupled QCD
as well as O.W.~Greenberg and V.~Vento for comments on the manuscript.
This work is supported by grant \#INV00--01--24 of La Conselleria de Cultura,
Educacio i Ciencia of the Val\`encian Community under the auspices of the program 
``Ayuda para Estancia de Investigadores Invitados en Universidades y Otros Centros 
de Investigacion Radicados en la Comunidad Val\`enciana''.
%

%
\vfill\eject
%
%
%
\centerline{FIGURE CAPTIONS}
\vskip 1cm
\noindent FIGURE~1. \hspace{0.1 cm}
The solution to the gap equation Eq.~(\ref{eq:gap}) to lowest order in 
$N_{\rm c}$ for $M_0 = 0$. (a) Dynamical quark mass $m$ for $r$ = 0 as a
function of the effective coupling constant $K$. The critical coupling
constant $K_C$ is approximately 0.73. (b) $K_C$ as a function of the Wilson
parameter $r$.
\vskip 0.75cm
\noindent FIGURE~2. \hspace{0.1 cm}
The solution to the gap equation Eq.~(\ref{eq:gap}) for $r$ = 1 and $M_0 = 0$. 
The coefficient of the momentum dependent dynamical 
quark mass $B$ from Eq.~(\ref{eq:gapq}) is shown
as a function of the effective coupling constant $K$. The critical coupling
constant $K_C$ is approximately 0.84.
\vfill\eject
\centerline{\Large\bf Figure 1}
\psfig{figure=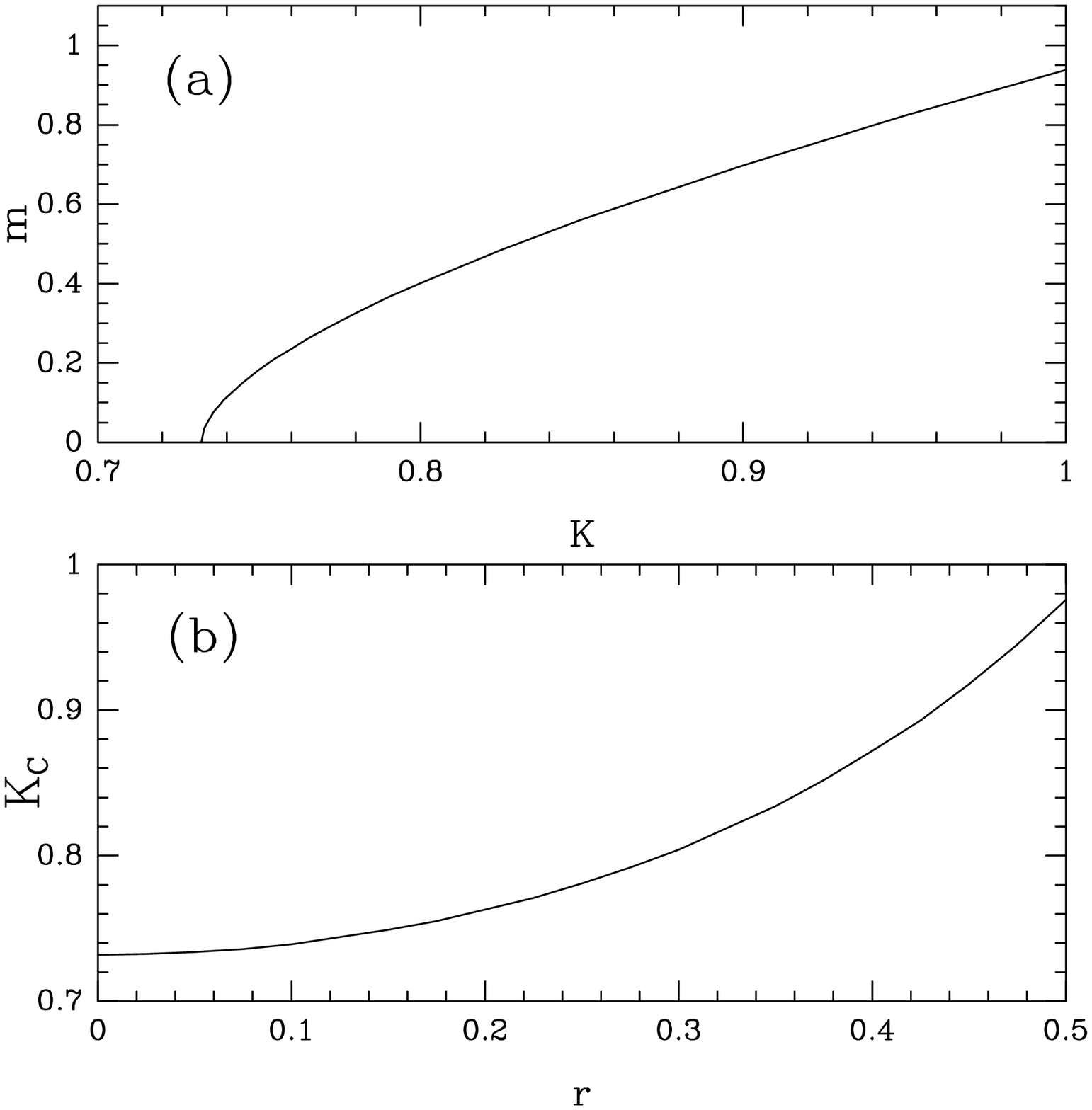,height=10cm,width=8cm}
\vfill\eject
\centerline{\Large\bf Figure 2}
\psfig{figure=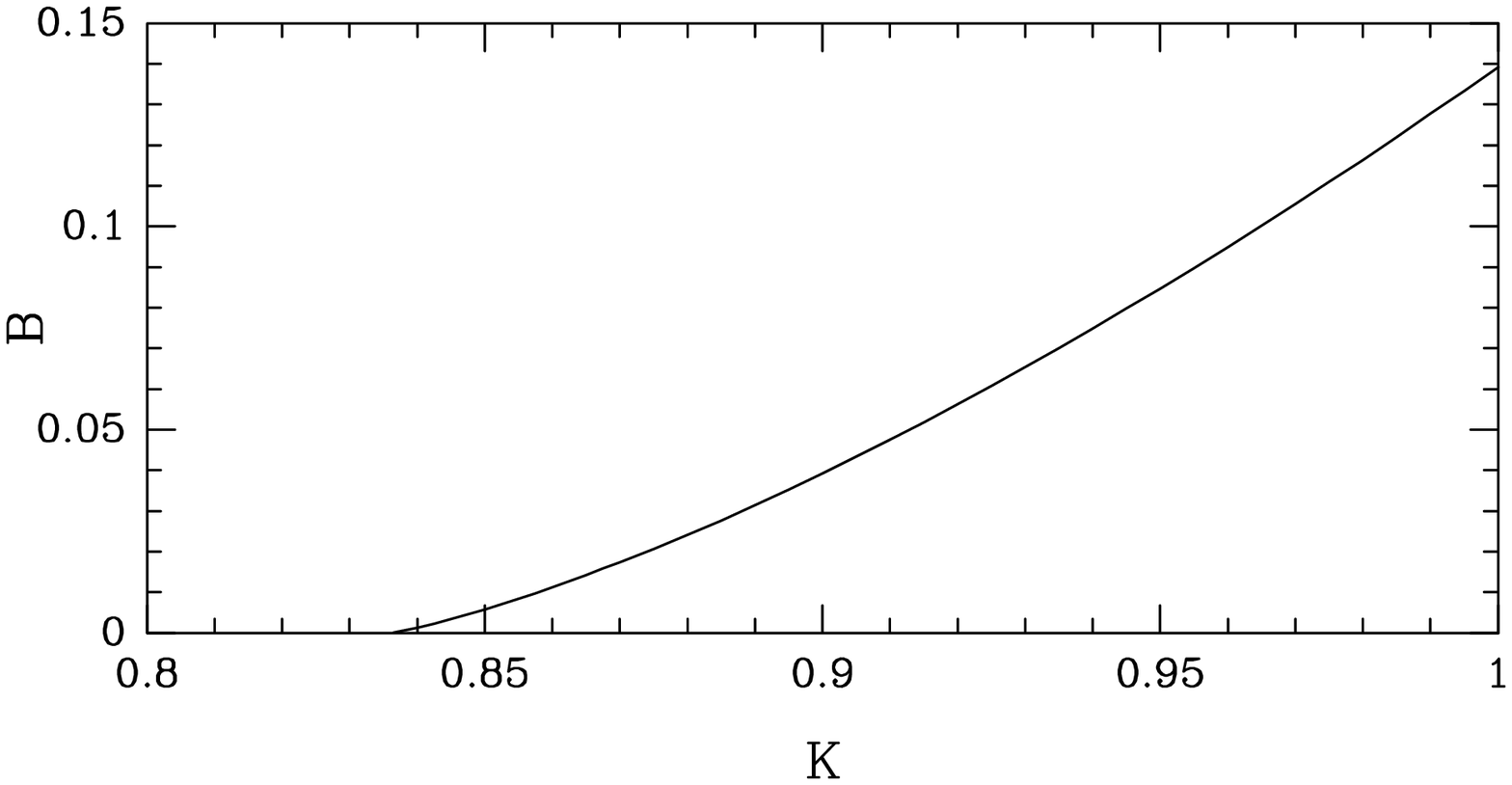,height=10cm,width=8cm}
\end{document}